\documentclass[aps,prl,twocolumn,showpacs,superscriptaddress,groupedaddress]{revtex4}  
\usepackage{graphicx}  
\usepackage{dcolumn}   
\usepackage{bm}        
\usepackage{amssymb}   
\usepackage{color}
\usepackage[colorlinks,]{hyperref}
\usepackage{mathrsfs}
\usepackage{amsfonts}
\usepackage{amsthm}
\usepackage{dcolumn}
\usepackage{txfonts}
\begin{document}

\widetext


\title{Protection of center-spin coherence by dynamically polarizing nuclear spin core in diamond}
\author {Gang-Qin Liu}
\thanks{These authors contribute equally}
\author {Qian-Qing Jiang}
\thanks{These authors contribute equally}
\author{Yan-Chun Chang}
\author{Dong-Qi Liu}
\author {Wu-Xia Li}
\author{Chang-Zhi Gu}
\affiliation{ Beijing National Laboratory for Condensed Matter
Physics, Institute of Physics, Chinese Academy of Sciences, Beijing
100190, China }
\author {Hoi Chun Po}
\affiliation{Department of Physics, The Chinese University of Hong
Kong, Shatin, New Territories, Hong Kong, China}
\author {Wen-Xian Zhang}
\affiliation{School of Physics and Technology, Wuhan University,
Wuhan, Hubei 430072, China}
\author{Nan Zhao}
\affiliation{Beijing Computational Science Research Center, Beijing
100084, China}
\author{Xin-Yu Pan$^{1}$}
\email{xypan@iphy.ac.cn}

\date{\today}

\begin{abstract}
We experimentally investigate the protection of electron spin
coherence of nitrogen vacancy (NV) center in diamond by dynamical
nuclear polarization. The electron spin decoherence of an NV center
is caused by the magnetic field fluctuation of the $^{13}$C nuclear
spin bath, which contributes large thermal fluctuation to the center
electron spin when it is in equilibrium state at room temperature.
To address this issue, we continuously transfer the angular momentum
from electron spin to nuclear spins, and pump the nuclear spin bath
to a polarized state under Hartman-Hahn condition. The bath
polarization effect is verified by the observation of prolongation
of the electron spin coherence time ($T_2^*$). Optimal conditions
for the dynamical nuclear polarization (DNP) process, including the
pumping pulse duration and depolarization effect of laser pulses,
are studied. Our experimental results provide strong support for
quantum information processing and quantum simulation using
polarized nuclear spin bath in solid state systems.
\end{abstract}

\pacs{76.70.Fz, 03.65.Yz,03.67.Lx,67.30.hj}
 \maketitle

Quantum information processing requires qubits that can be
initialized, controlled and readout with high fidelity. Furthermore,
the quantum coherence of qubits should persist for a long time to
realize multiple gate operations on them \cite{Nielsen book}.
However there exists inevitable noise from the environment that
causes decoherence of the qubits. Many efforts have been done to
protect the qubit from the noise.

Two major strategies have been proposed to enhance the coherent time
of qubits, namely, dynamical decoupling (DD)
\cite{Ban_DD}-\cite{YangW_DD} and dynamical nuclear polarization
(DNP)\cite{Reilly_DNP_Science}-\cite{ZhangWX_DNP}. DD can average
out the fluctuation of the spin bath by flipping center spin state,
thus could effectively cut off the interaction between the center
spin and its surrounding spin bath. For DNP method, in the ideal
case, the nuclear spin bath is prepared in a spin polarized state,
and the thermal fluctuation is completely suppressed. In this case,
the electron spin could have long coherence time even in the absence
of spin echo control (i.e. $T_2^* \sim T_2$).

DD can be used to protect the coherence of NV electron spin
\cite{Hanson_NV_DD},  \cite{Cory_NV_DD} but, in general, DD
sequences like CPMG or UDD, do not commute with quantum gate
operations, so one cannot realize gate operations on the protected
spin during DD sequence unless special designed pulses are applied
\cite{Van_DDwhileGate}, \cite {Du_CDD}. On the contrary, if one uses
DNP to generate a polarized nuclear spin bath, the center spin state
can be well protected and manipulated under the polarized
surrounding bath, leading to full exploitation of the center spin
coherence \cite{Reilly_DNP_Science}, \cite {DNP_Coherence_03PRL} and
\cite{DNP_Coherence_06PRL}. Furthermore, the small magnetic moment
of nucleus (3 orders smaller than electron spin) makes the bath
polarization persist for a very long time. The polarized nuclear
spins are important quantum resources for quantum information
processing and quantum simulation applications \cite{Cai2013NP}.

\begin{figure}
\includegraphics[width=\linewidth]{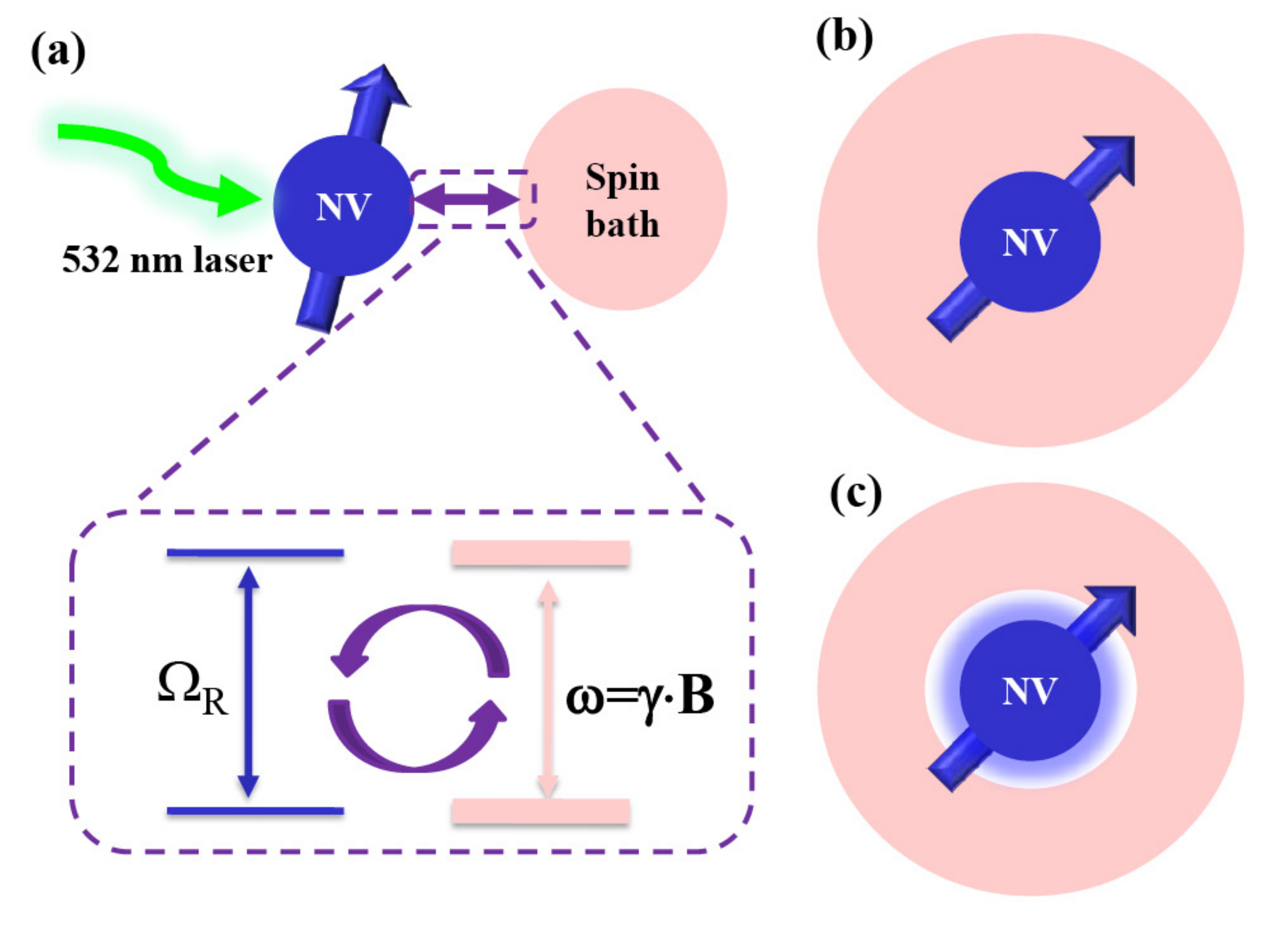} \caption{\label{fig:epsart}
(Color online) General Schematic. (a) Polarization transformation
around an NV center in diamond. Inset, the Hartman-Hahn condition
[see Eq.(1)]. (b) Spin bath without DNP, the pink shadow area
represents the unpolarized $^{13}$C nuclear spins in thermal
equilibrium. (c) After DNP process, the $^{13}$C nuclear spins close
to the NV center are polarized (the area in light blue).}
\end{figure}


Figure 1 shows the general idea of this paper. Firstly, a short
($\sim\mu$s) laser pulse is used to polarize the electron spin;
which serves as injection of spin polarization. Then, the electron
spin is brought to contact with the nuclear spin bath, so that the
angular momentum can transfer from the electron spin to the bath. An
efficient transfer channel is built up by driving the electron spin
with the Hartman-Hahn condition \cite{Hartmann_Hahn}(see below),
where the nuclear spins are resonant with the electron spin in the
rotating frame. The polarization of the bath spins is verified by
the enhancement of the electron spin coherence time. If the
surrounding bath spins are not polarized [Fig. 1b], the thermal
fluctuation of the bath spins induces a fast decoherence of the
central spin. However, when the nuclear spins in the surrounding
core are polarized [Fig. 1c], the thermal fluctuation is greatly
suppressed. The central spin is effectively isolated from the bath
spins, and the quantum coherence can persist for a longer time
\cite{ZhangWX_DNP}.

The key point for realizing the DNP is to build up an efficient
polarization transfer channel \cite{Hartmann_Hahn}. Generally
speaking, for two spins with different precession frequencies, the
angular momentum transfer process is inefficient due to the energy
mismatch. In order to overcome the energy mismatch, one can drive
the spins, and adjust the driving power (or Rabi frequency), so that
they are in resonance in the rotating frame
\cite{Belthangady_NVDNP_electron} \cite{Laraoui_NVDNP_electron}. In
our experiment, we drive the electron spin with microwave pulses,
and tune the Rabi frequency ($\Omega_R$) of electron spin to satisfy
the Hartman-Hahn condition.

\begin{equation}
\Omega_R=\gamma_C {\mathbf B}
\end{equation}

Where ${\mathbf B}$ is the magnetic field and
$\gamma_{\text{C}}=6.73\times 10^7$~T$^{-1}$s$^{-1}$ is the
gyromagnetic ratio of $^{13}$C. In this case, the Rabi frequency is
in resonance with the Larmor frequency of the $^{13}$C bath spins
under a certain external magnetic field [see the insert figure of
Fig. 1a]. Physically, the energy mismatch is compensated by the
microwave driving power, and the angular momentum transfer process
can occur efficiently.

\begin{figure}
\includegraphics[width=\linewidth]{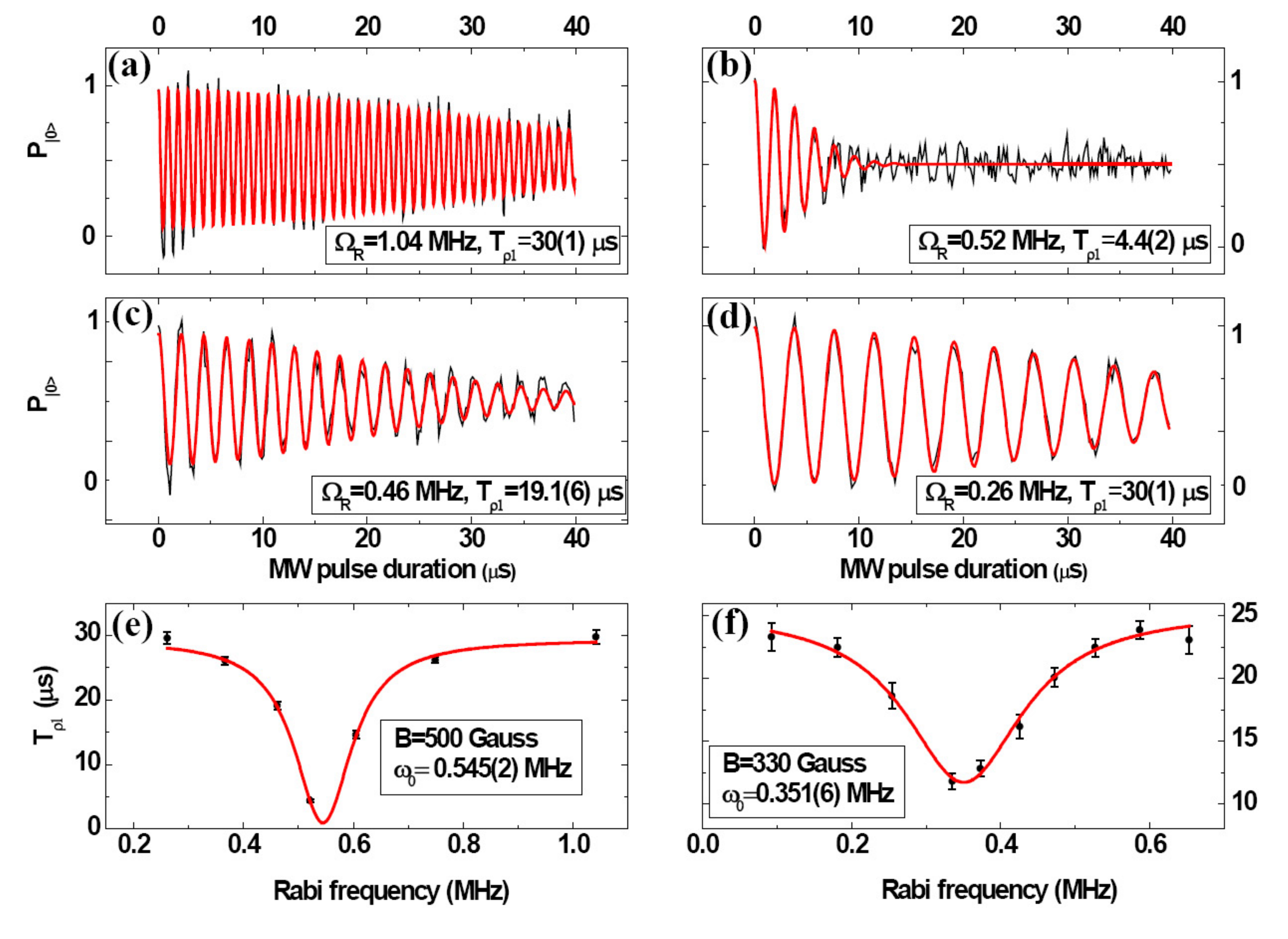} \caption{\label{fig:epsart}
(Color online) Hartman-Hahn resonance between electron spin and
nearby $^{13}$C nuclear spins. (a-d) Rabi oscillation of central
electron spin driven by different MW power under ${\mathbf
B}$=500~Gauss. (e) Dependence of the envelope decay time
(T$_{\rho1}$ time) on the Rabi frequency. A dramatic decay was
observed at the Hartman Hahn condition ($\Omega_R= \gamma_C {\mathbf
B}= 545$~kHz). (f) the same as e, with a different magnetic field
${\mathbf B}$= 330~Gauss ($\Omega_R= \gamma_C {\mathbf B}=
352$~kHz).}
\end{figure}

 We experimentally demonstrate this effect
by using an NV center in pure single crystal diamond (with Nitrogen
concentration $\ll$ 5~ppb). To enhance the photon collection
efficiency, a 12~$\mu$m diameter solid immersion lens (SIL) is
etched above the selected NV \cite{SIL_APL}. A coplanar waveguide
(CPW) antenna with 40~$\mu$m gap is deposited close to the SIL,
which can deliver microwave pulses to NV center with high
efficiency. A permanent magnet is used to generate the external
magnetic field ($\sim$ 10$^2$~Gauss) along [111] direction of the
crystal. The magnetic field lifts the degeneracy of the $m_S=\pm1$
states of the electron spin; on the other hand, the host $^{14}$N
nuclear spin is also polarized to the $m_I=+1$ sublevel by the
excited state level anti-crossing (ESLAC)\cite{ESLAC_PRB} under such
magnetic field. Thus we have a pure two-level system in the region
of weakly driven Rabi oscillation [See SI for details]. To
demonstrate the controllable transfer of spin polarization, we
change the power of microwave pulse, and measure the decay behavior
of Rabi oscillation. Figure 2 shows that the amplitude decay rate of
the Rabi oscillation strongly depends on the microwave power. In
particular, when the Hartman-Hahn condition is fulfilled, the
amplitude decay rate of Rabi oscillation is drastically increased.
While in the off-resonant regions, either $\Omega_R\gg\gamma_C
{\mathbf B}$ or $\Omega_R\ll\gamma_C {\mathbf B}$ , the Rabi
oscillation amplitude decays at much slower rate, indicating that
the polarized center spin is well isolated from the surrounding bath
spins. We fit the Rabi oscillation envelop, and extract the
characteristic decay time T$_{\rho1}$ . The dependence of decay time
T$_{\rho1}$ on Rabi frequency $\Omega_R$ under different magnetic
fields is summarized in Fig.2 e\& f). The decrease of T$_{\rho1}$
under Hartman-Hahn condition is caused by the resonant flip-flop
process between nuclear bath spins and the electron spin in the
rotating frame. While outside the resonant regime, the direct
flip-flop process is suppressed, and only the much weaker
second-order processes contribute to decay of the Rabi oscillation
amplitude \cite{Dobrovitski_RabiDecay}.


We now turn to the DNP of surrounding nuclear spin bath. Fig. 3a
shows the pulse sequence. A short 532 nm laser pulse (typically 3
$\mu$s) polarizes the center electron spin to $m_S=0$ state with
high fidelity (~95\% \cite{Van_DDwhileGate}).
 A  $\pi/2$ MW pulse rotates the state to the x direction in the equatorial plane of its Bloch sphere.
 A following 90-phase-shifted MW pulse with the same frequency locks the center spin state along
  the x direction in the rotating frame for a period of time $\tau$. The power of the spin-lock pulse is
  adjusted to match the HHDR condition, thus the optical pumped high polarization of center spin
  will transfer during the spin-lock pulse. After that, another laser pulse is used to polarize
  the center spin again. These polarization injection and transfer processes are repeated \emph{N} times,
  so that a significant nuclear spin polarization is built up in the bath. Finally, we check the
  polarization effect by measuring the Ramsey interference.

\begin{figure}
\includegraphics[width=\linewidth]{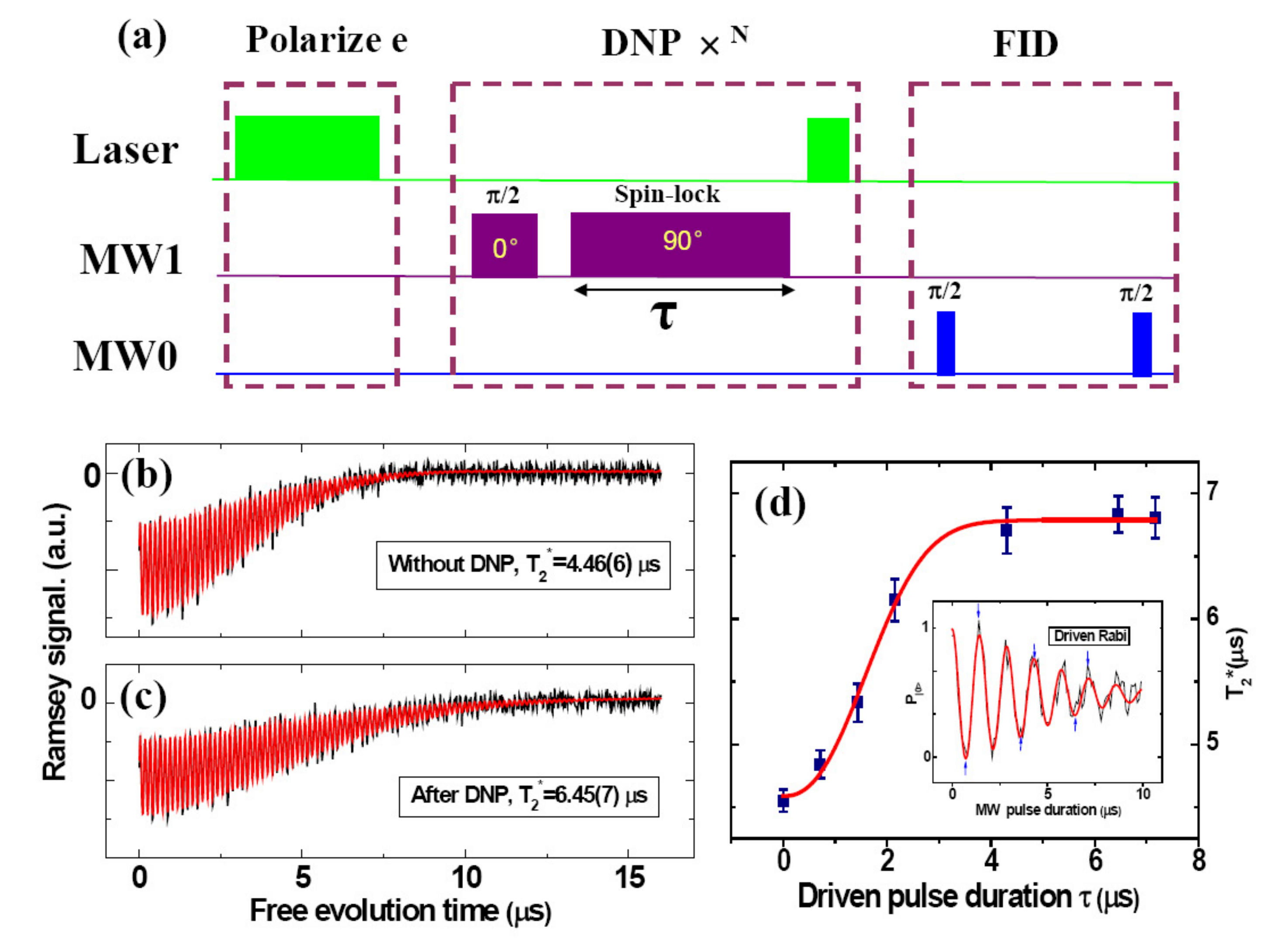} \caption{\label{fig:epsart}
(Color online) DNP pulse sequence and enhancement of $T_2^\star$
time. (a) The pulse sequence used to generate (the DNP sequence) and
to examine (the FID sequence) the nuclear spin bath polarization.
(b) FID without DNP.  (c) FID after DNP, the prolongation of the
dephasing time indicates that the bath spin is polarized during the
DNP process. (d) Dependence of the $T_2^\star$ time on the DNP
pumping duration ¦Ó, with \emph{N}=10 (Insert: driven Rabi). The
magnetic field is 660 Gauss for these measurements. }
\end{figure}

Fig. 3c shows the enhancement of $T_2^\star$ time under DNP. In an
external magnetic field of 660 Gauss, with \emph{N}= 10 times DNP
pumping pulses inserted before the Ramsey measurement sequence, the
electron spin coherence shows a Gaussian shape decay with a
characteristic decay time $T_2^\star= 6.45(7)~\mu s$. In comparison,
the Ramsey signal without DNP decays faster with $T_2^\star=
4.46(6)~\mu s$ (Fig. 3b). The fast oscillation is caused by the
strongly coupled nearby $^{13}$C nuclear spin. The enhancement of
the $T_2^\star$ time confirms that the DNP sequences suppress the
thermal fluctuation of the nuclear spin bath. Fig. 3d presents the
dependence of electron spin decoherence time on the spin-lock
duration $\tau$. We find a typical $2~\mu s$ lock time is sufficient
to transfer the center spin polarization, and longer lock time has
little improvement since $T_2^\star$  saturates at  $3~\mu s$. The
limitation of the depasing time is determined by the depolarizing
dynamics of the bath spins, including the influence of the laser
pulse, the fluctuation of the external magnetic field and so on.


\begin{figure}
\includegraphics[width=\linewidth]{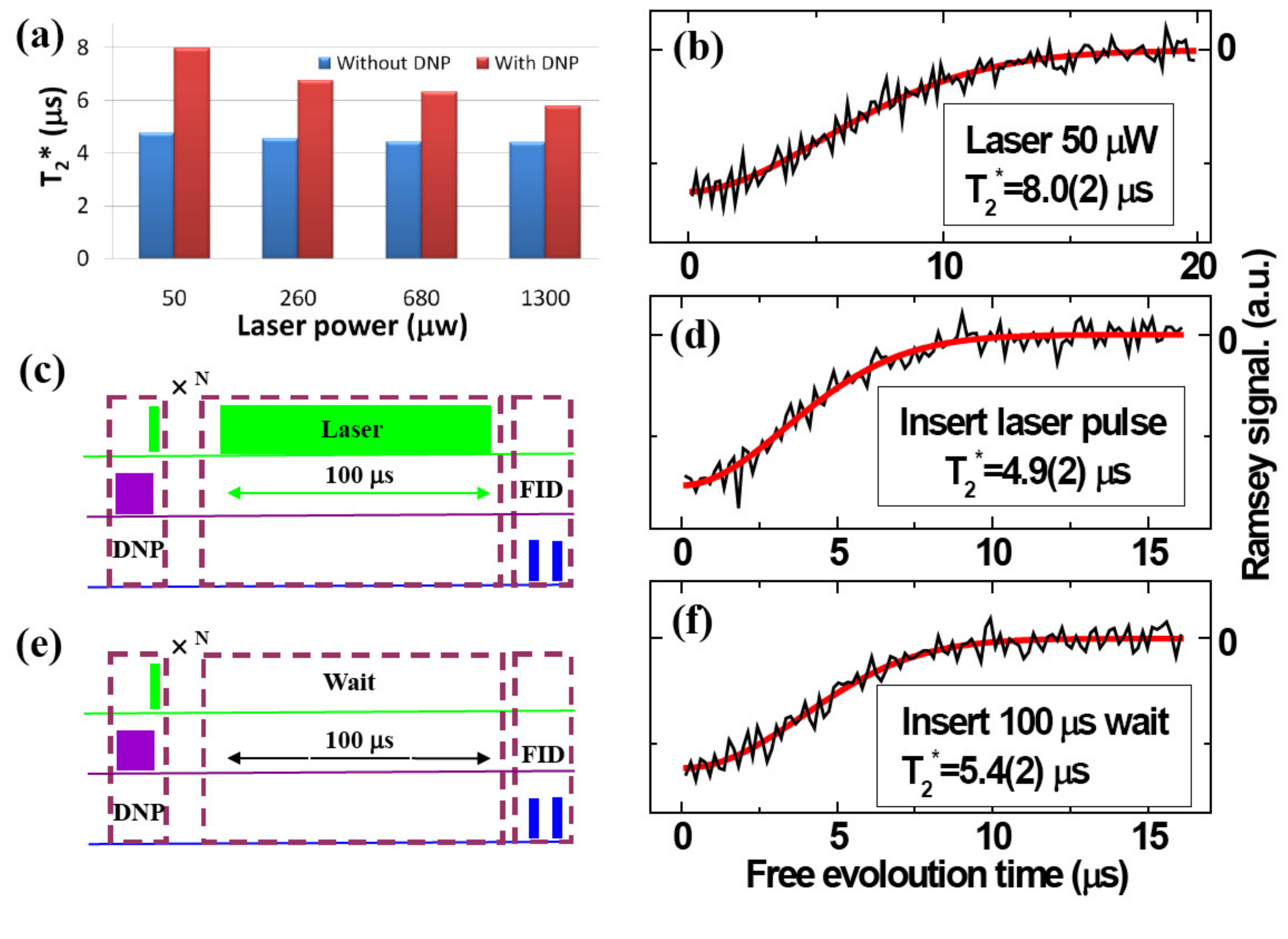} \caption{\label{fig:epsart}
 (Color online) Influence of laser pulse. (a) Laser power Dependence. The (red) blue columns
 are the $T_2^\star$ times measured with (without) DNP pumping process under
  4 different laser power.  (b) The FID result with DNP pumping at the
  optimized  laser power (50~$\mu$W). The $T_2^\star$ time of electron
   spin is $\sim$2 times longer than the case without DNP pumping.
    (c-d) Pulse sequence and result of adding depolarizing laser pulse.
    The insertion of laser pulses will depolarize the bath spin polarization
    built by DNP process, and result in a dephasing time close to the
     case without DNP. (e-f) Pulse sequence and result of adding waiting
     time during DNP process. The extra waiting time decreases the
     effective number of DNP circle, so an in-between dephasing time is observed.}
\end{figure}
We investigate the influence of laser power on DNP effect. The laser
pulses with power ranging from 50~$\mu$W to 1.3~mW were used to
carry out the same measurement sequence as shown in Fig. 3a. The
optical readout time for electron spin states is also adjusted to
ensure high readout fidelity in different laser power. For a fair
comparison, free induction decays under the same condition are also
measured and plotted [See SI for details.]. As depicted in Fig. 4a,
the enhancement of electron spin dephasing time are observed for all
the 4 measured cases, and lower laser power gives longer decoherence
time, which implies that the possible depolarization effect during
the excitation of the NV center electron. Fig. 4b shows the longest
coherence time we observed for this NV under the same magnetic field
after optimizing the laser power (50~$\mu$W was used). The observed
$T_2^\star= 8.0(2)~\mu$s is nearly two times compares to the
dephasing time without DNP.

To further characterize the influence of laser pulse, we insert an
extra laser pulse (100~$\mu$s) between the DNP and the standard FID
probe pulse (Fig. 4c). The measured result is presented by Fig. 4d,
the resultant dephasing time $T_2^\star= 4.9(2)~\mu$s  is just a
little longer than the dephasing time without DNP pulses (Fig. 3b),
indicates the nuclear spin bath polarization build by DNP process is
significantly destroyed by the laser pulse. A comparison pulse
sequence with the laser pulse replaced by a free period (Fig. 4e)
gives a long time $T_2^\star= 5.4(2)~\mu$s (see Fig. 4f) than the
case shown in Fig. 4d. Since the nuclear bath spins evolve in a time
scale of hundreds of microseconds, and the FID probe sequence takes
only several microseconds, the adjacent DNP interact with each other
in repetitive measurement, thus the effective number
\emph{N$^{\prime}$} for the continuous measurement is larger than
\emph{N}. When the 100~$\mu$s waiting pulse is inserted between the
DNP and FID sequence, the overlapped influence of adjacent DNP
sequences become weaker, and the effective number
\emph{N$^{\prime}$} decrease, thus the electron spin dephasing time
is shorter than that obtained with continuous measurement.


\emph{ Conclusions} We observed the polarization transfer between
center electron spin and $^{13}$C nuclear bath spins under
Hartman-Hahn condition, and investigated the $^{13}$C nuclear spin
dynamical polarization effect. Nearly 2 times longer center spin
dephasing time has been observed. The polarization of the spin bath
can be built in several microseconds in continuous measurement and
persist for hundreds of microseconds. We also found lower power
laser pulse could lead to gave longer dephasing time. These findings
could potentially open a new way to investigate the dynamics of spin
bath.

This work was supported by National Basic Research Program of China
(973 Program project Nos. 2009CB929103 and 2009CB930502), the
National Natural Science Foundation of China (Grants Nos. 10974251,
91123004, 11104334, 50825206 and 11275139).

 \emph{Authors' Note}: During the preparation of this manuscript, we got aware of a similar
investigation \cite{Jelezko DNP} in which the Hartmann-Hahn
condition was used to investigate the  $^{13}$C DNP in large
magnetic fields. We thank Fedor Jelezko for the helpful discussion.

\end{document}